\documentclass[aps,prl,twocolumn,groupedaddress,amsmath,amssymb,preprintnumbers]{revtex4}
\usepackage{color,slashed,graphicx,changes}
\usepackage[utf8]{inputenc}

\usepackage{hyperref}
\usepackage{soul,xcolor}
\usepackage{graphicx}
\usepackage{amsmath}
\usepackage{amssymb}
\usepackage{verbatim}
\usepackage{mathrsfs}
\usepackage{array}
\usepackage{layout}
\usepackage{textcomp}
\usepackage{latexsym}
\usepackage{slashed}
\usepackage{rotating}
\usepackage{hyperref}
\usepackage{booktabs}
\usepackage{relsize}
\usepackage{multirow}
\usepackage{bigints}

\newcommand\cyansout{\bgroup\markoverwith{\textcolor{cyan}{\rule[0.5ex]{2pt}{0.4pt}}}\ULon}
\newcommand\blueout{\bgroup\markoverwith{\textcolor{blue}{\rule[0.5ex]{2pt}{0.4pt}}}\ULon}

\newcommand{\Eq}{&=&}

\newcommand{\hs}[1]{{\hspace{#1}}}
\newcommand{\vs}[1]{{\vspace{#1}}}
\newcommand{\tf}[1]{{\textsf{#1}^{}}}
\newcommand{\tx}[1]{{\text{#1}^{}}}
\newcommand{\nn}{\nonumber\\}
\newcommand{\rx}[2]{{\raisebox{#1}{#2}}}
\newcommand{\sx}[2]{{\scalebox{#1}{#2}}}

\newcommand{\UoneX}{\tx{U}(1)_{\tf{L}_\mu - \tf{L}_\tau}^{}}
\newcommand{\BKnunu}{B^+ \!\to K^+\nu\bar{\nu}}

\begin{document}


\title{A Unified Dark Matter Explanation for $\boldsymbol{\BKnunu}$ \\ and the Super-Kamiokande Antineutrino Excess}



\author{Shu-Yu\,\,Ho$^{a}$} \thanks{shuyuho@as.edu.tw}

\author{Jongkuk\,\,Kim$^{b}$}
\thanks{jkkim@kangwon.ac.kr, {\it correspondence}}

\author{Pyungwon\,\,Ko$^{c}$}
\thanks{pko@kias.re.kr}

\affiliation{$^a$Institute of Physics, Academia Sinica, Nangang, Taipei 11529, Taiwan}
\affiliation{$^b$Division of Liberal Studies, Kangwon National University, Samcheok 25913, Korea}
\affiliation{$^c$Quantum Universe Center, KIAS, Seoul 02455, Korea}



\begin{abstract}
Recent results from Super-Kamiokande and Belle II have revealed intriguing excesses over Standard Model expectations.\,\,Super-Kamiokande observes a mild excess of $\bar{\nu}_e^{}$-like events near $20\,\,\mathrm{MeV}$, while Belle II reports a branching fraction for $\BKnunu$ that exceeds the Standard Model prediction by approximately $2.7\sigma$.\,\,In this work, we study the simplest UV-complete complex scalar dark matter model with a gauged $\UoneX$ symmetry.\,\,We demonstrate that a light dark sector can simultaneously reproduce the observed dark matter relic density and accommodate both excesses within a unified framework.

\end{abstract}

\pacs{}

\maketitle

\section{Introduction}
The Super-Kamiokande (SK) detector~\cite{Suzuki:2019jby} has recently begun operating in its gadolinium-doped configuration (SK-Gd), realized by dissolving gadolinium sulfate in the water target~\cite{Super-Kamiokande:2021the,Super-Kamiokande:2024kcb}.\,\,The idea, put forward in Ref.\,\cite{Beacom:2003nk}, exploits the fact that neutron capture on Gd is followed by a $\gamma$ cascade totalling $\sim\!\!8\,\,\mathrm{MeV}$, which provides a clean delayed tag for final-state neutrons.\,\,Such tagging sharply suppresses backgrounds and boosts the reach of SK for neutron-associated processes, foremost the diffuse supernova neutrino background (DSNB).

Super-Kamiokande has reported an excess of $\bar\nu_e^{}$ events around $E_{\bar\nu} \approx 22$~MeV, which lies within the primary search window for the diffuse supernova neutrino background (DSNB).\,\,This excess prefers the presence of a nonzero DSNB component over the background-only hypothesis with a significance of up to $2.3\,\sigma$, although theoretical uncertainties in the DSNB prediction leave room for alternative interpretations \cite{Super-Kamiokande:2021jaq,Super-Kamiokande:2025sxh}.\,\,
In Ref.\,\cite{Granelli:2026bem}, this mild excess observed by SK was interpreted as a possible signal of thermal dark matter (DM) in the tens of MeV mass range that annihilates predominantly into neutrinos through an $s$-wave process.\,\,Two scenarios were considered\,\,:\,\,direct annihilation into a neutrino-antineutrino pair and annihilation into a pair of dark-sector particles that subsequently decay into neutrinos.

For the direct-annihilation case, the preferred region is centered at $m_{\rm DM}^{} \simeq 22\,\,{\rm MeV}$ with
$\mathcal{J}_{\rm avg}^{} \langle\sigma v\rangle \in [1.8\times10^{-25},\,2.3\times10^{-24}]~{\rm cm^3\,s^{-1}}$
at $2\,\sigma$ C.L.~\cite{Granelli:2026bem, Endo:2026upb}.\,\,For the cascade-annihilation scenario, a similar fit is obtained for small mass splittings.\,\,As an illustrative example, the case of $\Delta = 0.03$ is considered, where $\Delta \equiv 1 \,-\, m_{\eta}/m_{\rm DM}^{}.$
Here, $\eta$ denotes the new dark-sector particle appearing in the final state.
The preferred parameters are
$m_{\rm DM}^{} \simeq 44\,\,{\rm MeV}$ and
$\mathcal{J}_{\rm avg}^{} \langle\sigma v\rangle \in [9.1\times10^{-26},\,1.1\times10^{-24}]~{\rm cm^3\,s^{-1}}$ at $2\,\sigma$ C.L.~\cite{Granelli:2026bem}.\,\,Here $\mathcal{J}_{\rm avg}^{}$ is the  the averaged J-factor normalized to $\mathcal{J}_0^{}$.\,\,In this work, we focus on the latter scenario, where DM annihilates into intermediate dark-sector states that subsequently decay into neutrino pairs.

In 2023, the Belle II Collaboration reported the first evidence for the rare decay $\BKnunu$~\cite{Belle-II:2023esi}.\,\,Using both hadronic-tagged and inclusive-tagged reconstruction techniques, the collaboration obtained the combined branching-fraction measurement
\begin{eqnarray}
{\cal B}(\BKnunu)_\mathrm{exp}^{} 
\,=\, 
(2.3 \pm 0.7) \times 10^{-5}
~.
\end{eqnarray}
From a theoretical perspective, $\BKnunu$ is one of the cleanest flavor-changing neutral-current observables.\,\,Owing to the small hadronic uncertainties and well-controlled electroweak corrections, its Standard Model (SM) prediction can be determined with remarkable precision~\cite{Parrott:2022zte},
\begin{eqnarray}
{\cal B}(\BKnunu)_\mathrm{SM}^{} 
\,=\, 
(4.97 \pm 0.37) \times 10^{-6}
~.
\end{eqnarray}
A comparison between the measured value and the SM prediction indicates an excess corresponding to a significance of approximately $2.7\,\sigma$.\,\,Although the current discrepancy is not statistically compelling enough to claim evidence for new physics, it nevertheless provides an intriguing motivation to explore possible extensions of the SM.\,\,In particular, scenarios involving light invisible particles or a dark sector offer a well-motivated framework in which the observed excess may be accommodated.

Motivated by these observations, we examine whether a minimal light dark sector can provide a common origin for both excesses.\,\,We consider a complex scalar DM model based on a gauged $\UoneX$ symmetry, in which DM annihilation generates the SK antineutrino signal while dark-Higgs-mediated processes contribute to the missing-energy events observed by Belle II.

\section{Model Overview}
A simple and well-motivated extension of the SM is the gauged $\UoneX$ framework, which is anomaly-free without introducing additional chiral fermions~\cite{He:1990pn,He:1991qd}.\,\,We extend this setup by introducing a dark Higgs field and a complex scalar DM candidate $X$.\,\,Within this framework, the excess observed in the $B^+\to K^+\nu\bar{\nu}$ channel can be interpreted as arising from invisible dark-sector particles, while the observed DM relic abundance is obtained through annihilation processes involving dark-sector states.

The DM particle is identified with a complex scalar field carrying a nonzero $\UoneX$ charge.\,\,We focus on the regime $m_X^{} > m_{Z'}^{}$, where DM annihilation is dominated by
\begin{eqnarray}
X X^\ast \rightarrow Z'Z'
~,
\end{eqnarray}
followed by subsequent decays of the dark gauge bosons $Z'$ predominantly into neutrinos.\,\,This differs from the conventional Higgs-portal singlet scalar DM scenario, in which DM annihilation typically proceeds into electrically charged SM particles~\cite{Silveira:1985rk,He:2009yd,He:2010nt}.

In this model, the relevant $\UoneX$ charge assignments are
\begin{eqnarray}
&&\widehat{{\cal Q}}_{\,\tf{L}_\mu - \tf{L}_\tau}^{}
\big(\nu_\mu,\,\nu_\tau,\,\mu,\,\tau,\,X,\,\Phi\big)
\nn
\Eq
\big(1,\,-1,\,1,\,-1,\,{\cal Q}_X^{},\,{\cal Q}_\Phi^{}\big)
~,
\end{eqnarray}
where
\begin{eqnarray}
\Phi
\,=\,
\frac{1}{\sqrt2} \big(\upsilon_\Phi^{} + \phi\big)
\end{eqnarray}
is the dark Higgs field.\,\,The vacuum expectation value $\upsilon_\Phi^{}$ breaks the $\UoneX$ symmetry spontaneously.\,\,To ensure the cosmological stability of DM, we take ${\cal Q}_X^{} = 1$ and choose ${\cal Q}_\Phi^{}$ such that no gauge-invariant operators up to dimension five induce DM decay~\cite{Baek:2013qwa,Ko:2022kvl}.

The renormalizable Lagrangian relevant for our analysis is given by
\begin{eqnarray}
{\cal L}
\Eq
|{\cal D}_{\rho}\Phi|^2
+
|{\cal D}_{\rho}X|^2
-
\tfrac14
Z'_{\rho\omega}
Z'^{\rho\omega}
-
m_X^2 |X|^2
\nn
&&
-\,\,
g_X^{}
\sx{1.2}{\big(}\,
\bar{\mu}\, \gamma^\rho\mu
-
\bar{\tau}\, \gamma^\rho\tau
+
\bar{\nu}_{\mu}^{} \gamma^\rho P_L^{} \nu_{\mu}^{}
-
\bar{\nu}_{\tau}^{} \gamma^\rho P_L^{} \nu_{\tau}^{}
\sx{1.2}{\big)}
Z'_\rho
\nn
&&
-\,\,
\lambda_{H X}^{}
|{\cal H}|^2_0 \,\,
|X|^2 
-
\lambda_{\Phi X}^{}
|\Phi|^2_0 \,\,
|X|^2 \,
-
\lambda_{H \Phi}^{} \,\,
|{\cal H}|^2_0 \,\,
|\Phi|^2_0
\nn[0.15cm]
&&
+\,\,\cdots
~,
\end{eqnarray}
where $Z'_{\rho\omega} = \partial_\rho Z'_\omega
- \partial_\omega Z'_\rho$, $P_L^{} = \tfrac12 (1 -\gamma^5)$, $|\Phi|^2_0 = |\Phi|^2- \upsilon_\Phi^2 / 2$, and $|{\cal H}|^2_0 = |{\cal H}|^2 -\upsilon_H^2/2$ with $\upsilon_H^{}$ being the vacuum expectation value of $H$.\,\,The covariant derivative is defined as
\begin{eqnarray}
{\cal D}_\rho^{}
\,=\,
\partial_\rho^{}
+
i\,g_X^{}
\widehat{{\cal Q}}_{\,\tf{L}_\mu - \tf{L}_\tau}^{}
Z'_\rho 
\end{eqnarray}
with $g_X^{}$ being the gauge coupling of $\UoneX$ symmetry, and the dark gauge boson (dark photon) mass is
\begin{eqnarray}
m_{Z'}^{}
\,=\,
g_X^{} |{\cal Q}_\Phi^{}| \,\upsilon_\Phi^{}
~.
\end{eqnarray}

The portal coupling $\lambda_{H \Phi}$ induces mixing between the CP-even scalar fields $h$ and $\phi\,$. The mass eigenstates are given by
\begin{eqnarray}
H_1 \,=\, \phi \,\, c_\theta^{} - h \, s_\theta^{}
~,\quad
H_2 \,=\, \phi \,\, s_\theta^{} + h \, c_\theta^{}
~,
\end{eqnarray}
where $H_2$ is identified with the observed $125\,\,{\rm GeV}$ Higgs boson, $c_\theta^{} = \cos\theta$ and $s_\theta^{} = \sin\theta$ with $\theta$ being the mixing angle, and $m_{H_2}^{} > m_{H_1}^{}$ is assumed.

Although absent at tree level, kinetic mixing between the dark photon and the ordinary photon is generated radiatively by $\mu$- and $\tau$-lepton loops, leading to an effective mixing parameter $\epsilon \simeq -\,\,g_X/70$~\cite{Escudero:2019gzq}.\,\,The DM phenomenology is mainly controlled by $g_X^{}$, $\lambda_{\Phi X}$, and the scalar spectrum.\,\,For simplicity, we set $\lambda_{H X}^{} = 0$ in the numerical analysis.\,\,The relevant parameter space is
\begin{eqnarray}
\big\{
m_{Z'}^{},~
m_X^{},~
m_{H_1}^{},~
\lambda_{\Phi X}^{},~
g_X^{},~
{\cal Q}_\Phi^{},~
\theta
\big\}
~.
\end{eqnarray}

\section{Super-Kamiokande $\bar{\nu}$ excess}
The DSNB provides a unique probe of the integrated history of stellar collapse in the Universe.\,\,Its expected flux depends on several astrophysical ingredients, including the cosmic core-collapse rate, neutrino emission from collapsing stars, and the fraction of failed supernovae.\,\,As a result, current theoretical predictions still carry sizable uncertainties \cite{Horiuchi:2008jz,Nakazato:2015rya,Kresse:2020nto,Tabrizi:2020vmo}.\,\,The recent gadolinium upgrade of SK significantly improves neutron tagging through the delayed gamma cascade following neutron capture on Gd, enhancing the sensitivity to inverse beta decay events.\,\,Inverse beta decay, $\bar{\nu}_e^{} + p\rightarrow e^+ + n$, provides the dominant detection channel for DSNB searches~\cite{Beacom:2003nk,Super-Kamiokande:2021jaq,Super-Kamiokande:2025sxh}.

When the full set of SK runs is analyzed jointly, the data exhibit a weak tendency toward a nonzero DSNB component, disfavoring the background-only hypothesis at about $2.3\,\sigma$.\,\,At present, it remains unclear whether this excess arises from the DSNB, a statistical fluctuation, or new physics.\,\,Motivated by this result, Ref.\,\cite{Granelli:2026bem} investigated a DM interpretation in which neutrinos are produced either directly from DM annihilation or via decays of intermediate dark-sector states.\,\,In the direct-annihilation scenario, the resulting neutrino spectrum is monochromatic, and a fit to the SK data favors a DM mass of
\begin{eqnarray}
m_{\rm DM}^{}
\,\simeq\, 22.1\,\,{\rm MeV}
~,
\end{eqnarray}
corresponding to a local significance of about $2.57\,\sigma$~\cite{Granelli:2026bem}.

In the $\UoneX$ scenario, the decay of $Z'$ produces neutrinos predominantly in the $\bar{\nu}_\mu^{}$ and $\bar{\nu}_\tau^{}$ flavors.\,\,Because the propagation length from the Galactic halo to the detector vastly exceeds the neutrino oscillation length, the interference terms wash out and the flux arrives at SK as an incoherent superposition of mass eigenstates.\,\,In this limit, the conversion probability into electron flavor is given by~\cite{Pakvasa:2007dc}
\begin{eqnarray}
P_{\alpha e}^{}
\,=\,
\sum_j \, |U_{\alpha j}^{}|^2 |U_{e j}^{}|^2
~,
\end{eqnarray}
where $U$ denotes the PMNS matrix.

In Ref.\,\cite{Endo:2026upb}, the fraction of the combined $\bar{\nu}_\mu^{}$ and $\bar{\nu}_\tau^{}$ flux observed as $\bar{\nu}_e$ at SK is defined as
\begin{eqnarray}
f_e^{}
\,=\,
\frac{P_{\mu e}^{} + P_{\tau e}^{}}{2}
\,=\, 
\frac{1 - P_{ee}^{}}{2}
\,\simeq\, 
0.225
~.
\label{eq:fe}
\end{eqnarray}
To account for the SK antineutrino excess, we consider $m_X^{}=\,44.4\,\,\mathrm{MeV}$ and
\begin{eqnarray}
&&\mathcal{J}_{\rm avg}^{}
\langle\sigma v\rangle_{XX^\ast \to Z'Z' \to 4\nu} 
\nn
&&
\in \left[ 2.0 \times 10^{-25},\, 2.41\times 10^{-24} \right]\,\,{\rm cm^3\,s^{-1}}
\end{eqnarray}
at $2\,\sigma$ C.L.\,\,For $\Delta = 1 -m_{Z'}^{}/m_X^{}= 0.03$, we take $m_{Z'}^{} \simeq 43\,\,\mathrm{MeV}$.\,\,In this mass range, $Z'$ decays predominantly into a pair of neutrinos.\,\,Here we note that profiling over the generalized NFW halo parameters yields a best-fit value $\mathcal{J}_{\rm avg} = 6^{+3}_{-1}$ at $1\sigma$, with $4 \lesssim \mathcal{J}_{\rm avg} \lesssim 17$ at $2\sigma$~\cite{Granelli:2026bem}.

\section{Higgs invisible decay}
Once the dark sector is coupled to the visible one, the $125\,\mathrm{GeV}$ state $H_2$ inherits new decay channels into light dark particles---$H_2 \to H_1 H_1$, $Z'Z'$, and $X X^\ast$---whenever phase space permits.\,\,These feed the (non-standard) invisible width of the Higgs, on which the LHC imposes ${\cal B}(H_2 \to \text{Inv.}) < 0.11$~\cite{ParticleDataGroup:2022pth}.\,\,The resulting bound on the dark sector, which we recap below following our earlier analysis~\cite{Ho:2024cwk}, mainly restricts the scalar mixing angle.

For $\lambda_{\Phi X} \lesssim \mathcal{O}(1)$, the exotic width is saturated by the $H_1H_1$ and $Z'Z'$ modes, with $\Gamma_{H_2\to H_1 H_1}\simeq \Gamma_{H_2 \to Z'Z'} \propto s_\theta^2\,\,m_{H_2}^3/\upsilon_\Phi^2$ exceeding $\Gamma_{H_2\to X X^\ast}\propto s_\theta^2\,\lambda_{\Phi X}^2\,\upsilon_\Phi^2/m_{H_2}^{}$.\,\,Because the two leading widths share the same $s_\theta^2/\upsilon_\Phi^2$ scaling, the invisible-decay limit collapses into a bound on the mixing angle alone, with little residual dependence on $\lambda_{\Phi X}$; numerically it demands $s_\theta^{} \lesssim 10^{-2}$.

The light $H_1$ bosons emitted in these decays could in principle reach SM fermions, but such modes are far subleading\,\,:\,\,the hierarchy $\Gamma_{H_1\to X X^\ast} \propto \lambda_{\Phi X}^2 \, \upsilon_\Phi^2 /m_{H_1}^{} \gg \Gamma_{H_1\to Z'Z'} \propto m_{H_1}^3/\upsilon_\Phi^2 \gg \Gamma_{H_1 \to f \bar{f}} \propto s_\theta^2\,m_f^2 m_{H_1}^{}/\upsilon_H^2$ holds throughout the region of interest.\,\,Consequently, every cascade terminates in invisible states, and in particular the visible chain $B^0 \to K^{\ast 0} H_1 \to K^{\ast 0}\mu^+\mu^-$ is rendered negligible~\cite{Ovchynnikov:2023von}.

\section{Dark Matter relic density}
Let us now discuss thermal freeze-out of the DM relic abundance.\,\,The DM number density evolves according to the Boltzmann equation.\,\,In order to obtain the observed relic abundance, the thermally averaged annihilation cross section for symmetric DM, $\Omega_X^{} = \Omega_{X^\ast}^{}$, satisfies~\cite{Griest:1990kh}
\begin{eqnarray}
\Omega_{\rm DM}^{} \hat{h}^2 
\,\simeq\, 
0.1 \times 
\frac{(20\,\,\mathrm{TeV})^{-2}}
{\langle \sigma v \rangle}
~.
\end{eqnarray}

If the dark Higgs is absent, a $\UoneX$-charged DM particle interacts with the SM only through $Z'$, so its annihilation rate is controlled by $m_{Z'}^{}$, $g_X^{}$, and $m_X^{}$ alone.\,\,With $m_{Z'}^{}$ and $g_X^{}$ already pinned by the $\Delta a_\mu$ constraint, reproducing the measured abundance then forces the narrow resonance window $m_{Z'}^{} \sim 2\,m_X^{}$.\,\,Introducing a dark Higgs relaxes this resonance condition, since it simultaneously supplies the $Z'$ mass and unlocks extra annihilation final states~\cite{Baek:2022ozm,Ho:2024cwk}.

With the dark Higgs included, the relic abundance is set by $X X^\ast \to Z'Z'$, supplemented by $H_1 Z'$ and $H_1 H_1$ once $m_X^{} \gtrsim m_{H_1}^{}$.\,\,Final states containing charged leptons or neutrinos are essentially irrelevant\,\,:\,\,$X X^\ast \to \ell^+\ell^-$ is penalized by $s_\theta^{}$, $X X^\ast \to \nu_\ell^{} \,\bar{\nu}_\ell^{}$ by $g_X^{}$, and the $e^+ e^-$ mode additionally by the tiny kinetic mixing.\,\,At small couplings the leading $X X^\ast \to Z'Z'$ amplitude is the $s$-channel $H_1$ exchange, with the pure-gauge pieces further suppressed by $g_X^4$.

Following the derivation in Ref.\,\cite{Ho:2024cwk}, the thermally averaged cross section in the $s_\theta^{} \ll 1$ limit takes the approximate form
\begin{eqnarray}
\langle \sigma v \rangle
\simeq
\frac{\lambda_{\Phi X}^2}{16\,\pi\,m_X^2}
\frac{\big(4 - 4\,r_{Z'}^2 + 3\,r_{Z'}^4\big)\sqrt{1 - r_{Z'}
^2}}
{\big(r_{H_1}^2 - 4\big)\rx{1pt}{$^{\!2}$} + r_{H_1}^2 \gamma_{H_1}^2 }
~,
\label{XX:ZpZp}
\end{eqnarray}
where $r_{Z'}^{} = m_{Z'}^{}/m_X^{}$, $r_{H_1}^{} = m_{H_1}^{}/m_X^{}$, and $\gamma_{H_1}^{} = \Gamma_{H_1}^{}/m_X^{}$ with $\Gamma_{H_1}^{}$ being the decay width of $H_1$,
\begin{eqnarray}
\Gamma_{H_1}^{}
\,\simeq\,
\frac{\lambda_{\Phi X}^2 \upsilon_\Phi^2}
{16\,\pi\,m_{H_1}^{}}
\sqrt{1 - \frac{4\,m_X^2}{m_{H_1}^2}}
~.
\end{eqnarray}
The competing $H_1 \to Z'Z'$ and $H_1 \to f\bar{f}$ widths are immaterial here, being throttled by the small $g_X^{}$ and $s_\theta^{}$.

A generic difficulty for light thermal DM arises from constraints on late-time energy injection into the primordial plasma~\cite{Slatyer:2015jla}.\,\,For conventional $s$-wave annihilation into visible SM particles, measurements of the cosmic microwave background (CMB) place strong limits on the annihilation cross section.\,\,In the present scenario, however, our setup evades the bound by a different route\,\,:\,\,although $X X^\ast \to Z'Z'$ and $H_1 H_1$ are $s$-wave, the daughters cascade as $H_1 \to X X^\ast$ and $Z' \to \nu \bar{\nu}$ (for $m_{Z'}^{} < m_\mu^{}$), so almost no energy is deposited into electromagnetic final states and the CMB limit is strongly diluted.\,\,The remaining handle is the shift in $N_{\rm eff}^{}$ from light DM annihilating to neutrinos, which disfavors a complex scalar below $8.2\,\,\mathrm{MeV}$~\cite{Chu:2023jyb}; our benchmark $m_X^{} \simeq 44\,\,\mathrm{MeV}$, fixed by the SK antineutrino excess, lies safely above this threshold.

\section{Two- or three-body decays at Belle II}
If interpreted as a signal of new physics, the Belle II excess points to an extra contribution to the $b \to s \nu \bar{\nu}$ transition that manifests as missing energy $\slashed{E}$ recoiling against the kaon in $B^+\to K^+$.\,\,Subtracting the SM piece, the new-physics branching fraction is
\begin{eqnarray}
{\cal B}\big(B^+ \to K^+ \slashed{E}\big)_{\rm NP}^{}
\,=\, 
(1.8 \pm 0.7) \times 10^{-5}
~.
\end{eqnarray}

A variety of theoretical interpretations have been proposed, ranging from model-independent effective field theory analyses to ultraviolet-complete constructions.\,\,In particular, shifts in the relevant Wilson coefficients have been widely studied~\cite{Athron:2023hmz,Bause:2023mfe,Allwicher:2023xba,He:2023bnk,Hou:2024vyw}.\,\,Alternatively, new invisible particles produced in three-body decays such as $B^+\to K^+\chi\bar{\chi}$ have been considered, where $\chi$ denotes a light hidden-sector state~\cite{Berezhnoy:2023rxx,Datta:2023iln,Altmannshofer:2023hkn,McKeen:2023uzo,Fridell:2023ssf,Cheung:2024oxh,Ho:2024cwk,Gabrielli:2024wys,Berezhnoy:2025nmb,Hu:2025zua}.\,\,A broad range of new-physics scenarios has also been explored in this context~\cite{Felkl:2023ayn,Wang:2023trd,He:2024iju,Bolton:2024egx,Rosauro-Alcaraz:2024mvx,Kim:2024tsm,Hati:2024ppg,Buras:2024ewl,Altmannshofer:2024kxb,Hu:2024mgf,Altmannshofer:2025eor,Calibbi:2025rpx,Lee:2025jky,He:2025jfc,Berezhnoy:2025tiw,Bolton:2025fsq,Aliev:2025hyp,Chen:2025npb,Ding:2025eqq,He:2025sao,DiLuzio:2025qkc,Shaw:2025ays,Berezhnoy:2025osn,Lee:2025kvf,Kim:2025zaf}.\,\,Among these possibilities, a light dark-sector particle contributing to $B^+\to K^+ + \slashed{E}$ provides a particularly appealing explanation of the missing-energy signature.

The reconstructed $q^2$ distribution carries further information\,\,:\,\,Ref.\,\cite{Altmannshofer:2023hkn} pointed to a localized bump near $q^2_{\rm rec}\simeq 4\,\,{\rm GeV}^2$, naturally reproduced by a $\sim\!2\,\mathrm{GeV}$ boson whose couplings to SM fermions are tiny.\,\,A Higgs-portal realization with three-body decay into a DM pair was instead studied in Ref.\,\cite{McKeen:2023uzo}, although it tends to overclose the Universe unless extra annihilation modes or a non-standard cosmology are added.\,\,The likelihood study of Ref.\,\cite{Fridell:2023ssf} moreover favors the three-body interpretation.\,\,Taken together, these studies suggest that light dark-sector states can provide a viable explanation for the excess.

The first bounds of this type trace back to Ref.\,\cite{Bird:2004ts}, which used $B^+ \to K^+ \nu \bar{\nu}$ to constrain a real-singlet scalar DM coupled through the Higgs portal.\,\,That candidate is nonetheless heavily disfavored by the CMB, since its abundance is fixed by the $s$-wave chain $S S \to H_2^{(*)}\to f \bar{f}$ into visible states.

By contrast, the dark Higgs of our model opens new $B^+$ decay topologies.\,\,When $m_{B^+}^{} - m_{K^+}^{} > m_{H_1}^{}$, the two-body mode $B^+\to K^+ H_1$ is kinematically open, with width, as obtained in our earlier work~\cite{Ho:2024cwk},
\begin{eqnarray}
\Gamma_{B^+ \to K^+ H_1}^{}
&\hs{-0.2cm}\simeq\hs{-0.2cm}&
\frac{|\kappa_{cb}^{}|^2 s_\theta^2}
{64\,\pi\,m_{B^+}^3}
\big[f_0^{}(m_{H_1}^2)\big]^2
\bigg(\frac{m_{B^+}^2 - m_{K^+}^2}{m_b^{} - m_s^{}}\bigg)^{\hs{-0.1cm}2}
\nonumber\\
&&\times
\sqrt{\mathcal{K}\big(m_{B^+}^2,\,m_{K^+}^2,\,m_{H_1}^2\big)}
~,
\end{eqnarray}
where $\kappa_{cb}^{} \simeq 6.7 \times 10^{-6}$ is the loop-induced effective coupling, $f_0^{}(q^2)$ is the $B\to K$ form factor~\cite{Parrott:2022rgu}, and $\mathcal{K}(a,b,c) = a^2 + b^2 + c^2 - 2 (a\,b + b\,c + c\,a)$.

When the dark Higgs is off-shell, i.e. $m_{H_1}^{} > m_{B^+}^{} - m_{K^+}^{} > 2\,m_X^{}$, the transition instead proceeds through the three-body chain $B^+ \to K^+ H_1^{(*)} \to K^+ X X^\ast$, whose width reads~\cite{Ho:2024cwk}
\begin{eqnarray}
\hs{-0.6cm}
\Gamma_{B^+ \to K^+ X X^\ast}
&\hs{-0.2cm}\simeq\hs{-0.2cm}&
\frac{\lambda_{\Phi X}^2 \upsilon_\Phi^2 
|\kappa_{cb}^{}|^2 s_\theta^2}
{1024\,\pi^3 m_{B^+}^3}
\bigg(\frac{m_{B^+}^2-m_{K^+}^2}{m_b^{} - m_s^{}}\bigg)^{\hs{-0.1cm}2}
\nonumber\\
&&\times
\bigintssss \!\! dt \,\,
\frac{\mathcal{I}(t)\,[f_0(t)]^2\,(m_{H_1}^2-m_{H_2}^2)^2}
{(t-m_{H_1}^2)^2 (t-m_{H_2}^2)^2}
~,
\end{eqnarray}
with $4\,m_X^2 \le t = q^2 \le (m_{B^+} ^{} - m_{K^+}^{})^2$ and
\begin{eqnarray}
\mathcal{I}(q^2)
\,=\,
\sqrt{1-\frac{4\,m_X^2}{q^2}}
\sqrt{\mathcal{K}(m_{B^+}^2,\,m_{K^+}^2,\,q^2)}
~.
\end{eqnarray}
Although the decay $B^+ \to K^+ Z'Z'$ is also kinematically allowed, its contribution is negligible since $\Gamma_{B^+ \to K^+ Z'Z'}/\Gamma_{B^+ \to K^+ X X^\ast} \propto m_B^4/(\lambda_{\Phi X}^2 \upsilon_\Phi^4)\lesssim 7 \times10^{-3}$ for $\lambda_{\Phi X}\gtrsim \mathcal{O}(0.1)$ required by relic density and both the Belle II and SK anti-neutrino excesses.

\section{Numerical analysis}
\begin{figure}
\centering
\includegraphics[width=0.98\linewidth]{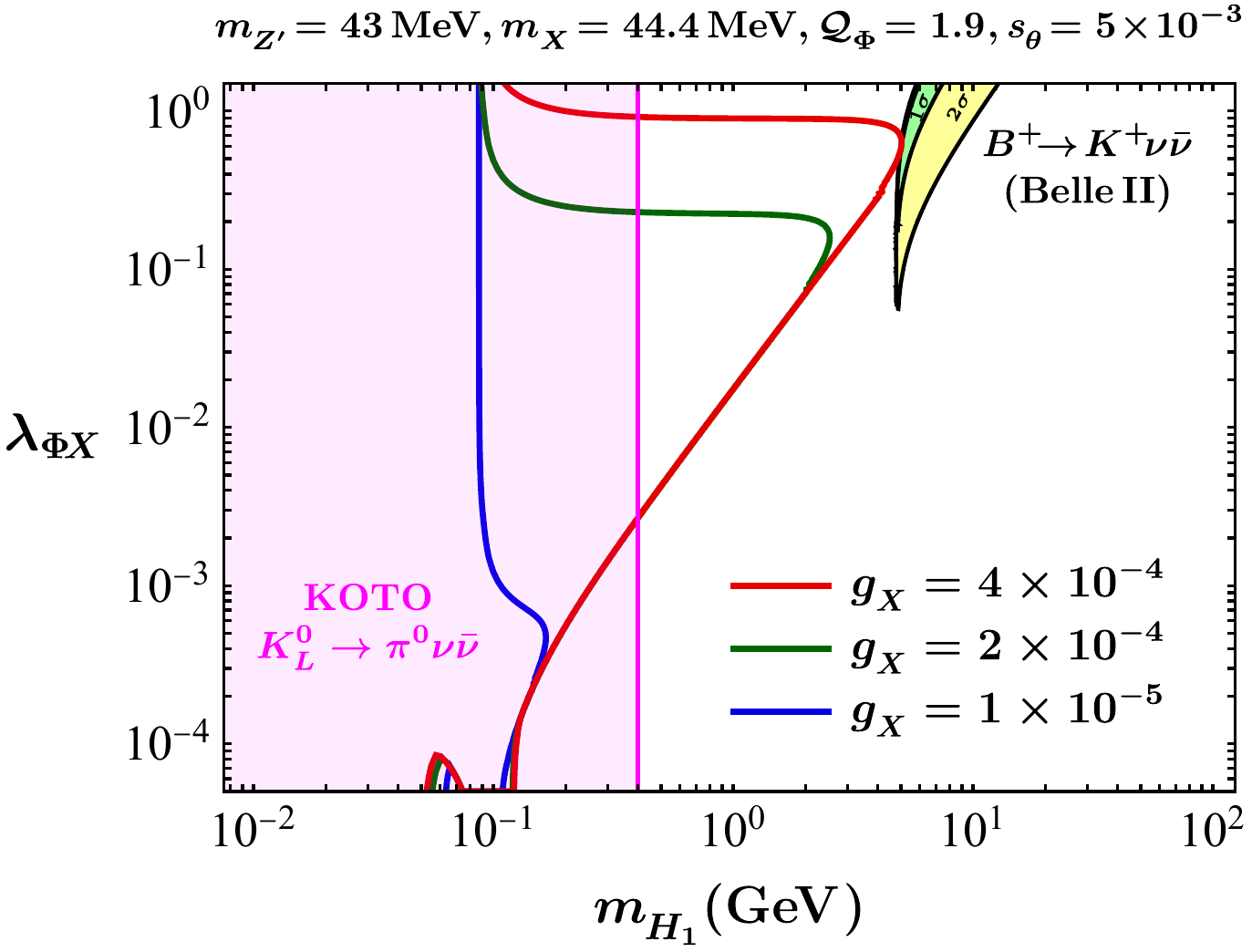}
\includegraphics[width=0.98\linewidth]{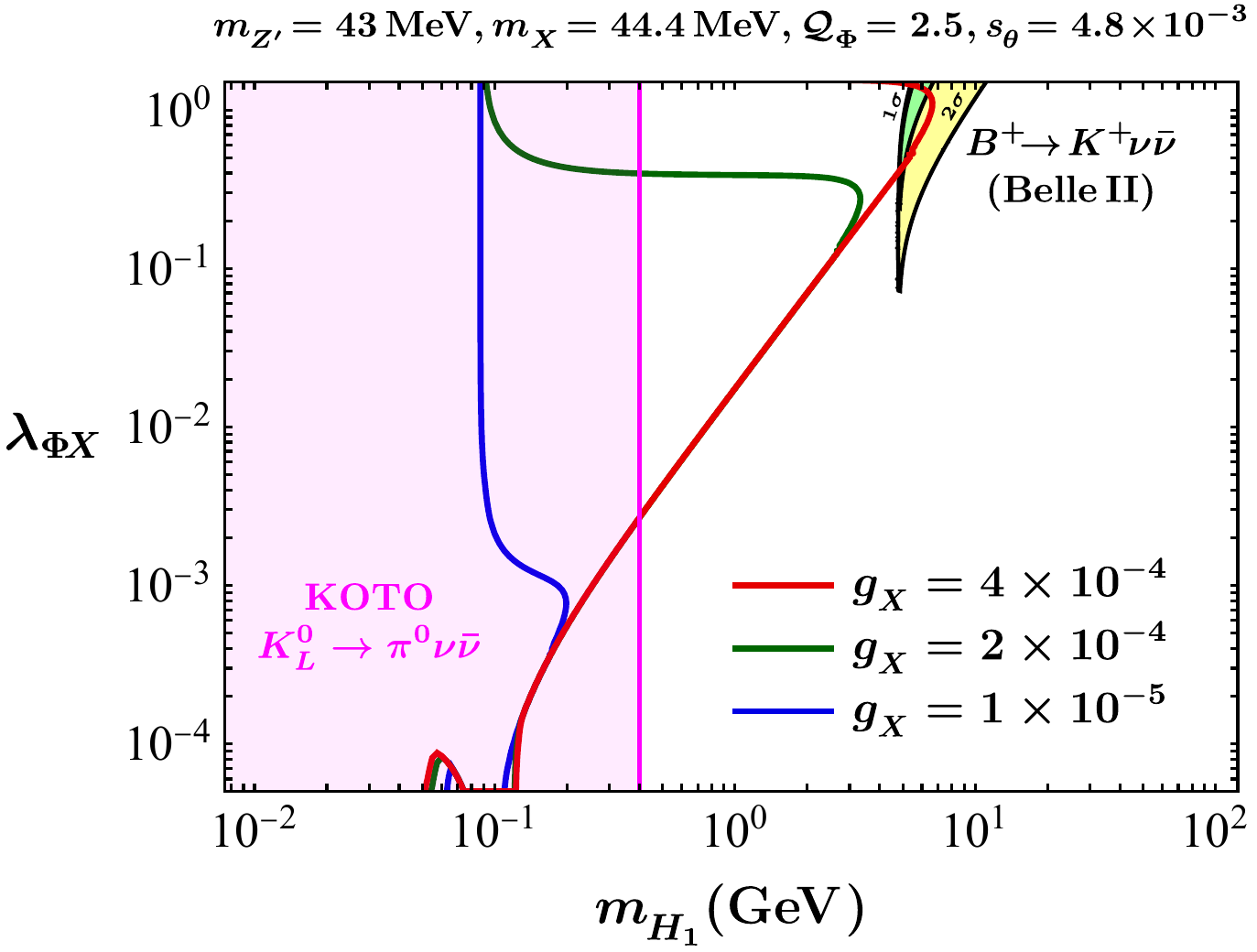}
\vs{-0.3cm}
\caption{Viable parameter space in the $(m_{H_1}^{},\,\lambda_{\Phi X}^{})$ plane for $m_{Z'}^{} = 43\,\,\mathrm{MeV}$ and $m_X^{} = 44.4\,\,\mathrm{MeV}$, with ${\cal Q}_\Phi^{} = 1.9$, $s_\theta^{} = 5 \times 10^{-3}$ (upper) and ${\cal Q}_\Phi^{} = 2.5$, $s_\theta^{} = 4.8 \times 10^{-3}$ (bottom).\,\,The colored curves provide the observed relic abundance for different values of the gauge coupling, $g_X^{} = 4 \times 10^{-4}$ (red), $g_X^{} = 2 \times 10^{-4}$ (green), and $g_X^{} = 10^{-5}$ (blue), respectively.\,\,When ${\cal Q}_\Phi^{} \gtrsim 1.9$ and $g_X^{} \simeq 4 \times 10^{-4}$, the Belle II excess can be explained through the three-body decay channel.} 
\label{fig:relic}
\vs{-0.3cm}
\end{figure}

In Fig.\,\ref{fig:relic}, we show the regions of the $(m_{H_1}^{},\,\lambda_{\Phi X}^{})$ parameter space that are consistent with all relevant requirements for the two representative values ${\cal Q}_\Phi^{} = 1.9$ (upper) and ${\cal Q}_\Phi^{} = 2.5$ (bottom).\,\,We fixed $m_{Z'}^{} = 43\,\,\mathrm{MeV}$ and $m_X^{} = 44.4\,\,\mathrm{MeV}$ to explain the SK anti-$\nu$ excess.\,\,Each colored curve corresponds to the correct relic density $\Omega_{\rm DM}^{}\hat{h}^{2}\simeq 0.12$ for different values of the gauge coupling, $g_X^{} = 4 \times 10^{-4}$ (red), $g_X^{} = 2 \times 10^{-4}$ (green), and $g_X^{} = 10^{-5}$ (blue), respectively.\,\,After the dark Higgs resonance, the $X X^{*}\to Z'Z'$ annihilation channel becomes dominant.\,\,The dip around $m_{H_1}^{} \simeq 2\,m_X^{}$ happens because of the $s$-channel $H_1$ resonance at $m_{H_1}^{} \simeq 2\,m_X^{}$.\,\,We find that for $\lambda_{\Phi X}^{} \gtrsim \mathcal{O}(0.1)$ the model accommodates the relic density together with both excesses across a broad span of $m_{H_1}^{}$, with the small mixing $s_\theta^{} \simeq 5 \times 10^{-3}$ ensuring consistency with the invisible-Higgs decay constraint.
Considering the constraint of the Higgs invisible decay width, choosing $\lambda_{\Phi X} \ll 1$ safely evades current constraints.

\section{Conclusions}
In this work, we have studied the simplest UV-complete complex scalar dark matter model based on a gauged $\UoneX$ symmetry, extended by a dark Higgs field.
And, we have shown that it provides a unified explanation of two recent excesses.\,\,A complex scalar dark matter candidate $X$ with $m_X^{} \simeq 44.4\,\,\mathrm{MeV}$ annihilates dominantly through the $s$-wave process $X X^{*}\to Z'Z'$, with the light dark photon ($m_{Z'}^{}\simeq 43\,\,\mathrm{MeV}$) decaying predominantly into a pair of either $\nu_\mu^{}$ or $\nu_\tau^{}$.\,\,After oscillations average out over astrophysical baselines, a fraction $f_e^{} \simeq 0.225$ of this flux is detected as $\bar\nu_e$, accommodating the Super-Kamiokande excess for $J_{\rm avg}\langle\sigma v\rangle\in [2.0 \times 10^{-25},\,2.41\times 10^{-24}]\,\mathrm{cm^{3}\,s^{-1}}$. 
Since the final states are neutrinos, the otherwise stringent CMB bounds on $s$-wave light dark matter are significantly relaxed \cite{Ho:2024cwk}.

The same dark sector accounts for the Belle~II excess through the two-body decay $B^+ \to K^+ H_1$, or the three-body decay $B^+ \to K^+ X X^{*}$ proceeding through an on- or off-shell dark Higgs.\,\,$H_1$ decays invisibly and the signal appears as missing energy consistent with $\mathcal{B}(B^+ \to K^+ +\slashed{E})_{\rm NP} \simeq1.8\times10^{-5}$.\,\,Consistency with the Higgs invisible bound from LHC requires $s_\theta^{}\lesssim 10^{-2}$, and as shown in Fig.\,\ref{fig:relic} the model produces the relic density while accommodating both excesses over a broad range of $m_{H_1}^{}$ and $\lambda_{\Phi X}^{}$.\,\,Although this excess is statistically mild, a single anomaly-free light dark sector can unify them, making this scenario an economical and testable target for future SK-Gd and Belle~II measurements.
\section{acknowledgments}
This study was supported by 2026 Research Grant from Kangwon National University.\,\,This work is supported by National Research Foundation of Korea (NRF) Research Grant No. RS-2024-00341419 (JK) and RS-2025-24803289 (PK).

\bibliography{main}

@article{Suzuki:2019jby,
    author = "Suzuki, Yoichiro",
    title = "{The Super-Kamiokande experiment}",
    doi = "10.1140/epjc/s10052-019-6796-2",
    journal = "Eur. Phys. J. C",
    volume = "79",
    number = "4",
    pages = "298",
    year = "2019"
}

@article{Super-Kamiokande:2021the,
    author = "Abe, K. and others",
    collaboration = "Super-Kamiokande",
    title = "{First gadolinium loading to Super-Kamiokande}",
    eprint = "2109.00360",
    archivePrefix = "arXiv",
    primaryClass = "physics.ins-det",
    doi = "10.1016/j.nima.2021.166248",
    journal = "Nucl. Instrum. Meth. A",
    volume = "1027",
    pages = "166248",
    year = "2022"
}

@article{Super-Kamiokande:2024kcb,
    author = "Abe, K. and others",
    collaboration = "Super-Kamiokande",
    title = "{Second gadolinium loading to Super-Kamiokande}",
    eprint = "2403.07796",
    archivePrefix = "arXiv",
    primaryClass = "physics.ins-det",
    doi = "10.1016/j.nima.2024.169480",
    journal = "Nucl. Instrum. Meth. A",
    volume = "1065",
    pages = "169480",
    year = "2024"
}

@article{Beacom:2003nk,
    author = "Beacom, John F. and Vagins, Mark R.",
    title = "{GADZOOKS! Antineutrino Spectroscopy with Large Water Cherenkov Detectors}",
    eprint = "hep-ph/0309300",
    archivePrefix = "arXiv",
    reportNumber = "FERMILAB-PUB-03-249-A",
    doi = "10.1103/PhysRevLett.93.171101",
    journal = "Phys. Rev. Lett.",
    volume = "93",
    pages = "171101",
    year = "2004"
}

@article{Belle-II:2023esi,
    author = "Adachi, I. and others",
    collaboration = "Belle-II",
    title = "{Evidence for B+{\textrightarrow}K+{\ensuremath{\nu}}{\ensuremath{\nu}}{\textasciimacron} decays}",
    eprint = "2311.14647",
    archivePrefix = "arXiv",
    primaryClass = "hep-ex",
    reportNumber = "Belle II Preprint 2023-017, KEK Preprint 2023-35",
    doi = "10.1103/PhysRevD.109.112006",
    journal = "Phys. Rev. D",
    volume = "109",
    number = "11",
    pages = "112006",
    year = "2024"
}

@article{Parrott:2022zte,
    author = "Parrott, W. G. and Bouchard, C. and Davies, C. T. H.",
    collaboration = "HPQCD",
    title = "{Standard Model predictions for B{\textrightarrow}K{\ensuremath{\ell}}+{\ensuremath{\ell}}-, B{\textrightarrow}K{\ensuremath{\ell}}1-{\ensuremath{\ell}}2+ and B{\textrightarrow}K{\ensuremath{\nu}}{\ensuremath{\nu}}{\textasciimacron} using form factors from Nf=2+1+1 lattice QCD}",
    eprint = "2207.13371",
    archivePrefix = "arXiv",
    primaryClass = "hep-ph",
    doi = "10.1103/PhysRevD.107.014511",
    journal = "Phys. Rev. D",
    volume = "107",
    number = "1",
    pages = "014511",
    year = "2023",
    note = "[Erratum: Phys.Rev.D 107, 119903 (2023)]"
}

@article{He:1990pn,
    author = "He, X. G. and Joshi, Girish C. and Lew, H. and Volkas, R. R.",
    title = "{NEW Z-prime PHENOMENOLOGY}",
    reportNumber = "UM-P-90/42, OZ-P-90/16",
    doi = "10.1103/PhysRevD.43.R22",
    journal = "Phys. Rev. D",
    volume = "43",
    pages = "22--24",
    year = "1991"
}

@article{He:1991qd,
    author = "He, Xiao-Gang and Joshi, Girish C. and Lew, H. and Volkas, R. R.",
    title = "{Simplest Z-prime model}",
    reportNumber = "CERN-TH-6084-91, UM-P-91-32, OZ-91-07",
    doi = "10.1103/PhysRevD.44.2118",
    journal = "Phys. Rev. D",
    volume = "44",
    pages = "2118--2132",
    year = "1991"
}

@article{Silveira:1985rk,
    author = "Silveira, Vanda and Zee, A.",
    title = "{SCALAR PHANTOMS}",
    reportNumber = "DOE-ER-40048-13 P5",
    doi = "10.1016/0370-2693(85)90624-0",
    journal = "Phys. Lett. B",
    volume = "161",
    pages = "136--140",
    year = "1985"
}

@article{He:2009yd,
    author = "He, Xiao-Gang and Li, Tong and Li, Xue-Qian and Tandean, Jusak and Tsai, Ho-Chin",
    title = "{The Simplest Dark-Matter Model, CDMS II Results, and Higgs Detection at LHC}",
    eprint = "0912.4722",
    archivePrefix = "arXiv",
    primaryClass = "hep-ph",
    doi = "10.1016/j.physletb.2010.04.026",
    journal = "Phys. Lett. B",
    volume = "688",
    pages = "332--336",
    year = "2010"
}

@article{He:2010nt,
    author = "He, Xiao-Gang and Ho, Shu-Yu and Tandean, Jusak and Tsai, Ho-Chin",
    title = "{Scalar Dark Matter and Standard Model with Four Generations}",
    eprint = "1004.3464",
    archivePrefix = "arXiv",
    primaryClass = "hep-ph",
    doi = "10.1103/PhysRevD.82.035016",
    journal = "Phys. Rev. D",
    volume = "82",
    pages = "035016",
    year = "2010"
}

@article{Baek:2013qwa,
    author = "Baek, Seungwon and Ko, P. and Park, Wan-Il",
    title = "{Singlet Portal Extensions of the Standard Seesaw Models to a Dark Sector with Local Dark Symmetry}",
    eprint = "1303.4280",
    archivePrefix = "arXiv",
    primaryClass = "hep-ph",
    doi = "10.1007/JHEP07(2013)013",
    journal = "JHEP",
    volume = "07",
    pages = "013",
    year = "2013"
}

@article{Ko:2022kvl,
    author = "Ko, P. and Lu, Chih-Ting and Min, Ui",
    title = "{Crossing two-component dark matter models and implications for 511 keV $\gamma$-ray and XENON1T excesses}",
    eprint = "2202.12648",
    archivePrefix = "arXiv",
    primaryClass = "hep-ph",
    month = "2",
    year = "2022"
}

@article{Escudero:2019gzq,
    author = "Escudero, Miguel and Hooper, Dan and Krnjaic, Gordan and Pierre, Mathias",
    title = "{Cosmology with A Very Light L$_{\mu}$ {\ensuremath{-}} L$_{\tau}$ Gauge Boson}",
    eprint = "1901.02010",
    archivePrefix = "arXiv",
    primaryClass = "hep-ph",
    reportNumber = "FERMILAB-PUB-19-001-A, LPT-Orsay-18-15, IFIC-19-02, KCL-19-01,
  IFT-UAM/CSIC-19-7, KCL-19-01",
    doi = "10.1007/JHEP03(2019)071",
    journal = "JHEP",
    volume = "03",
    pages = "071",
    year = "2019"
}

@article{Horiuchi:2008jz,
    author = "Horiuchi, Shunsaku and Beacom, John F. and Dwek, Eli",
    title = "{The Diffuse Supernova Neutrino Background is detectable in Super-Kamiokande}",
    eprint = "0812.3157",
    archivePrefix = "arXiv",
    primaryClass = "astro-ph",
    doi = "10.1103/PhysRevD.79.083013",
    journal = "Phys. Rev. D",
    volume = "79",
    pages = "083013",
    year = "2009"
}

@article{Nakazato:2015rya,
    author = "Nakazato, Ken'ichiro and Mochida, Eri and Niino, Yuu and Suzuki, Hideyuki",
    title = "{Spectrum of the Supernova Relic Neutrino Background and Metallicity Evolution of Galaxies}",
    eprint = "1503.01236",
    archivePrefix = "arXiv",
    primaryClass = "astro-ph.HE",
    doi = "10.1088/0004-637X/804/1/75",
    journal = "Astrophys. J.",
    volume = "804",
    number = "1",
    pages = "75",
    year = "2015"
}

@article{Kresse:2020nto,
    author = "Kresse, Daniel and Ertl, Thomas and Janka, Hans-Thomas",
    title = "{Stellar Collapse Diversity and the Diffuse Supernova Neutrino Background}",
    eprint = "2010.04728",
    archivePrefix = "arXiv",
    primaryClass = "astro-ph.HE",
    doi = "10.3847/1538-4357/abd54e",
    journal = "Astrophys. J.",
    volume = "909",
    number = "2",
    pages = "169",
    year = "2021"
}

@article{Tabrizi:2020vmo,
    author = "Tabrizi, Zahra and Horiuchi, Shunsaku",
    title = "{Flavor Triangle of the Diffuse Supernova Neutrino Background}",
    eprint = "2011.10933",
    archivePrefix = "arXiv",
    primaryClass = "hep-ph",
    doi = "10.1088/1475-7516/2021/05/011",
    journal = "JCAP",
    volume = "05",
    pages = "011",
    year = "2021"
}

@article{Super-Kamiokande:2021jaq,
    author = "Abe, K. and others",
    collaboration = "Super-Kamiokande",
    title = "{Diffuse supernova neutrino background search at Super-Kamiokande}",
    eprint = "2109.11174",
    archivePrefix = "arXiv",
    primaryClass = "astro-ph.HE",
    doi = "10.1103/PhysRevD.104.122002",
    journal = "Phys. Rev. D",
    volume = "104",
    number = "12",
    pages = "122002",
    year = "2021"
}

@article{Granelli:2026bem,
    author = "Granelli, Alessandro and Pascoli, Silvia and Rosauro-Alcaraz, Salvador",
    title = "{Dark Matter Interpretation of the Super-Kamiokande Antineutrino Excess and Predictions for JUNO}",
    eprint = "2605.20162",
    archivePrefix = "arXiv",
    primaryClass = "hep-ph",
    month = "5",
    year = "2026"
}

@article{Pakvasa:2007dc,
    author = "Pakvasa, Sandip and Rodejohann, Werner and Weiler, Thomas J.",
    title = "{Flavor Ratios of Astrophysical Neutrinos: Implications for Precision Measurements}",
    eprint = "0711.4517",
    archivePrefix = "arXiv",
    primaryClass = "hep-ph",
    doi = "10.1088/1126-6708/2008/02/005",
    journal = "JHEP",
    volume = "02",
    pages = "005",
    year = "2008"
}

@article{ParticleDataGroup:2022pth,
    author = "Workman, R. L. and others",
    collaboration = "Particle Data Group",
    title = "{Review of Particle Physics}",
    doi = "10.1093/ptep/ptac097",
    journal = "PTEP",
    volume = "2022",
    pages = "083C01",
    year = "2022"
}

@article{Ovchynnikov:2023von,
    author = "Ovchynnikov, Maksym and Schmidt, Michael A. and Schwetz, Thomas",
    title = "{Complementarity of $B\rightarrow K^{(*)} \mu \bar{\mu }$ and $B\rightarrow K^{(*)} + \textrm{inv}$ for searches of GeV-scale Higgs-like scalars}",
    eprint = "2306.09508",
    archivePrefix = "arXiv",
    primaryClass = "hep-ph",
    reportNumber = "CPPC-2023-02",
    doi = "10.1140/epjc/s10052-023-11975-0",
    journal = "Eur. Phys. J. C",
    volume = "83",
    number = "9",
    pages = "791",
    year = "2023"
}

@article{Griest:1990kh,
    author = "Griest, Kim and Seckel, David",
    title = "{Three exceptions in the calculation of relic abundances}",
    reportNumber = "CFPA-TH-90-001A, BA-90-79",
    doi = "10.1103/PhysRevD.43.3191",
    journal = "Phys. Rev. D",
    volume = "43",
    pages = "3191--3203",
    year = "1991"
}

@article{Baek:2022ozm,
    author = "Baek, Seungwon and Kim, Jongkuk and Ko, P.",
    title = "{Muon (g {\ensuremath{-}} 2) and thermal WIMP DM in $ \textrm{U}{(1)}_{L_{\mu }-{L}_{\tau }} $ models}",
    eprint = "2204.04889",
    archivePrefix = "arXiv",
    primaryClass = "hep-ph",
    reportNumber = "KIAS-P22019",
    doi = "10.1007/JHEP01(2025)014",
    journal = "JHEP",
    volume = "01",
    pages = "014",
    year = "2025"
}

@article{Slatyer:2015jla,
    author = "Slatyer, Tracy R.",
    title = "{Indirect dark matter signatures in the cosmic dark ages. I. Generalizing the bound on s-wave dark matter annihilation from Planck results}",
    eprint = "1506.03811",
    archivePrefix = "arXiv",
    primaryClass = "hep-ph",
    reportNumber = "MIT-CTP-4682",
    doi = "10.1103/PhysRevD.93.023527",
    journal = "Phys. Rev. D",
    volume = "93",
    number = "2",
    pages = "023527",
    year = "2016"
}

@article{Chu:2023jyb,
    author = "Chu, Xiaoyong and Pradler, Josef",
    title = "{Minimal mass of thermal dark matter and the viability of millicharged particles affecting 21-cm cosmology}",
    eprint = "2310.06611",
    archivePrefix = "arXiv",
    primaryClass = "hep-ph",
    doi = "10.1103/PhysRevD.109.103510",
    journal = "Phys. Rev. D",
    volume = "109",
    number = "10",
    pages = "103510",
    year = "2024"
}

@article{Athron:2023hmz,
    author = "Athron, Peter and Martinez, R. and Sierra, Cristian",
    title = "{B meson anomalies and large $ {B}^{+}\to {K}^{+}\nu \overline{\nu} $ in non-universal U(1)$^{′}$ models}",
    eprint = "2308.13426",
    archivePrefix = "arXiv",
    primaryClass = "hep-ph",
    doi = "10.1007/JHEP02(2024)121",
    journal = "JHEP",
    volume = "02",
    pages = "121",
    year = "2024"
}

@article{Bause:2023mfe,
    author = "Bause, Rigo and Gisbert, Hector and Hiller, Gudrun",
    title = "{Implications of an enhanced B{\textrightarrow}K{\ensuremath{\nu}}{\ensuremath{\nu}}{\textasciimacron} branching ratio}",
    eprint = "2309.00075",
    archivePrefix = "arXiv",
    primaryClass = "hep-ph",
    doi = "10.1103/PhysRevD.109.015006",
    journal = "Phys. Rev. D",
    volume = "109",
    number = "1",
    pages = "015006",
    year = "2024"
}

@article{Allwicher:2023xba,
    author = "Allwicher, Lukas and Becirevic, Damir and Piazza, Gioacchino and Rosauro-Alcaraz, Salvador and Sumensari, Olcyr",
    title = "{Understanding the first measurement of B(B{\textrightarrow}K{\ensuremath{\nu}}{\ensuremath{\nu}}{\textasciimacron})}",
    eprint = "2309.02246",
    archivePrefix = "arXiv",
    primaryClass = "hep-ph",
    doi = "10.1016/j.physletb.2023.138411",
    journal = "Phys. Lett. B",
    volume = "848",
    pages = "138411",
    year = "2024"
}

@article{He:2023bnk,
    author = "He, Xiao-Gang and Ma, Xiao-Dong and Valencia, German",
    title = "{Revisiting models that enhance B+{\textrightarrow}K+{\ensuremath{\nu}}{\ensuremath{\nu}}{\textasciimacron} in light of the new Belle II measurement}",
    eprint = "2309.12741",
    archivePrefix = "arXiv",
    primaryClass = "hep-ph",
    doi = "10.1103/PhysRevD.109.075019",
    journal = "Phys. Rev. D",
    volume = "109",
    number = "7",
    pages = "075019",
    year = "2024"
}

@article{Hou:2024vyw,
    author = "Hou, Biao-Feng and Li, Xin-Qiang and Shen, Meng and Yang, Ya-Dong and Yuan, Xing-Bo",
    title = "{Deciphering the Belle II data on $ B\to K\nu \overline{\nu} $ decay in the (dark) SMEFT with minimal flavour violation}",
    eprint = "2402.19208",
    archivePrefix = "arXiv",
    primaryClass = "hep-ph",
    doi = "10.1007/JHEP06(2024)172",
    journal = "JHEP",
    volume = "06",
    pages = "172",
    year = "2024"
}

@article{Berezhnoy:2023rxx,
    author = "Berezhnoy, Alexander and Melikhov, Dmitri",
    title = "{$B\to K^* M_X$ vs $B\to K M_X$ as a probe of a scalar-mediator dark matter scenario}",
    eprint = "2309.17191",
    archivePrefix = "arXiv",
    primaryClass = "hep-ph",
    doi = "10.1209/0295-5075/ad1d03",
    journal = "EPL",
    volume = "145",
    number = "1",
    pages = "14001",
    year = "2024"
}

@article{Datta:2023iln,
    author = "Datta, Alakabha and Marfatia, Danny and Mukherjee, Lopamudra",
    title = "{B{\textrightarrow}K{\ensuremath{\nu}}{\ensuremath{\nu}}{\textasciimacron}, MiniBooNE and muon g-2 anomalies from a dark sector}",
    eprint = "2310.15136",
    archivePrefix = "arXiv",
    primaryClass = "hep-ph",
    doi = "10.1103/PhysRevD.109.L031701",
    journal = "Phys. Rev. D",
    volume = "109",
    number = "3",
    pages = "L031701",
    year = "2024"
}

@article{Altmannshofer:2023hkn,
    author = "Altmannshofer, Wolfgang and Crivellin, Andreas and Haigh, Huw and Inguglia, Gianluca and Martin Camalich, Jorge",
    title = "{Light new physics in B{\textrightarrow}K(*){\ensuremath{\nu}}{\ensuremath{\nu}}{\textasciimacron}?}",
    eprint = "2311.14629",
    archivePrefix = "arXiv",
    primaryClass = "hep-ph",
    reportNumber = "PSI-PR-23-46, ZU-TH 77/23",
    doi = "10.1103/PhysRevD.109.075008",
    journal = "Phys. Rev. D",
    volume = "109",
    number = "7",
    pages = "075008",
    year = "2024"
}

@article{McKeen:2023uzo,
    author = "McKeen, David and Ng, John N. and Tuckler, Douglas",
    title = "{Higgs portal interpretation of the Belle II B+{\textrightarrow}K+{\ensuremath{\nu}}{\ensuremath{\nu}} measurement}",
    eprint = "2312.00982",
    archivePrefix = "arXiv",
    primaryClass = "hep-ph",
    doi = "10.1103/PhysRevD.109.075006",
    journal = "Phys. Rev. D",
    volume = "109",
    number = "7",
    pages = "075006",
    year = "2024"
}

@article{Fridell:2023ssf,
    author = "Fridell, K{\r{a}}re and Ghosh, Mitrajyoti and Okui, Takemichi and Tobioka, Kohsaku",
    title = "{Decoding the B{\textrightarrow}K{\ensuremath{\nu}}{\ensuremath{\nu}} excess at Belle II: Kinematics, operators, and masses}",
    eprint = "2312.12507",
    archivePrefix = "arXiv",
    primaryClass = "hep-ph",
    reportNumber = "KEK-TH-2587",
    doi = "10.1103/PhysRevD.109.115006",
    journal = "Phys. Rev. D",
    volume = "109",
    number = "11",
    pages = "115006",
    year = "2024"
}

@article{Cheung:2024oxh,
    author = "Cheung, Kingman and Kim, Yongkyu and Kwon, Youngjoon and Ouseph, C. J. and Soffer, Abner and Wang, Zeren Simon",
    title = "{Probing dark photons from a light scalar at Belle II}",
    eprint = "2401.03168",
    archivePrefix = "arXiv",
    primaryClass = "hep-ph",
    doi = "10.1007/JHEP05(2024)094",
    journal = "JHEP",
    volume = "05",
    pages = "094",
    year = "2024"
}

@article{Ho:2024cwk,
    author = "Ho, Shu-Yu and Kim, Jongkuk and Ko, Pyungwon",
    title = "{Recent B+{\textrightarrow}K+{\ensuremath{\nu}}{\ensuremath{\nu}}{\textasciimacron} excess and muon g-2 illuminating light dark sector with Higgs portal}",
    eprint = "2401.10112",
    archivePrefix = "arXiv",
    primaryClass = "hep-ph",
    reportNumber = "KIAS-24003, KIAS-p24003",
    doi = "10.1103/PhysRevD.111.055029",
    journal = "Phys. Rev. D",
    volume = "111",
    number = "5",
    pages = "055029",
    year = "2025"
}

@article{Gabrielli:2024wys,
    author = {Gabrielli, Emidio and Marzola, Luca and M{\"u}{\"u}rsepp, Kristjan and Raidal, Martti},
    title = "{Explaining the $B^+\rightarrow K^+ \nu \bar{\nu }$ excess via a massless dark photon}",
    eprint = "2402.05901",
    archivePrefix = "arXiv",
    primaryClass = "hep-ph",
    doi = "10.1140/epjc/s10052-024-12818-2",
    journal = "Eur. Phys. J. C",
    volume = "84",
    number = "5",
    pages = "460",
    year = "2024"
}

@article{Berezhnoy:2025nmb,
    author = "Berezhnoy, Alexander and Lucha, Wolfgang and Melikhov, Dmitri",
    title = "{Probing vector- versus scalar-mediator dark-matter scenarios in $B\rightarrow (K,K^*) M_X$ decays}",
    eprint = "2507.10801",
    archivePrefix = "arXiv",
    primaryClass = "hep-ph",
    doi = "10.1140/epjp/s13360-026-07673-x",
    journal = "Eur. Phys. J. Plus",
    volume = "141",
    number = "5",
    pages = "495",
    year = "2026"
}

@article{Hu:2025zua,
    author = "Hu, Quan-Yi and Duan, Zhi-Bin",
    title = "{Influence of invisible light particles on {\ensuremath{\Lambda}}b{\textrightarrow}{\ensuremath{\Lambda}}Emiss}",
    eprint = "2510.05272",
    archivePrefix = "arXiv",
    primaryClass = "hep-ph",
    doi = "10.1103/zp29-r1l6",
    journal = "Phys. Rev. D",
    volume = "112",
    number = "9",
    pages = "095007",
    year = "2025"
}

@article{Felkl:2023ayn,
    author = "Felkl, Tobias and Giri, Anjan and Mohanta, Rukmani and Schmidt, Michael A.",
    title = "{When energy goes missing: new physics in $b\rightarrow s \nu \nu $ with sterile neutrinos}",
    eprint = "2309.02940",
    archivePrefix = "arXiv",
    primaryClass = "hep-ph",
    reportNumber = "CPPC-2023-06",
    doi = "10.1140/epjc/s10052-023-12326-9",
    journal = "Eur. Phys. J. C",
    volume = "83",
    number = "12",
    pages = "1135",
    year = "2023"
}

@article{Wang:2023trd,
    author = {Wang, Zeren Simon and Dreiner, Herbert K. and G{\"u}nther, Julian Y.},
    title = "{The decay $B\rightarrow K+\nu +\bar{\nu }$ at Belle II and a massless bino in R-parity-violating supersymmetry}",
    eprint = "2309.03727",
    archivePrefix = "arXiv",
    primaryClass = "hep-ph",
    doi = "10.1140/epjc/s10052-025-13745-6",
    journal = "Eur. Phys. J. C",
    volume = "85",
    number = "1",
    pages = "66",
    year = "2025"
}

@article{He:2024iju,
    author = "He, Xiao-Gang and Ma, Xiao-Dong and Schmidt, Michael A. and Valencia, German and Volkas, Raymond R.",
    title = "{Scalar dark matter explanation of the excess in the Belle II B$^{+}${\textrightarrow} K$^{+}$+ invisible measurement}",
    eprint = "2403.12485",
    archivePrefix = "arXiv",
    primaryClass = "hep-ph",
    reportNumber = "CPPC-2024-04",
    doi = "10.1007/JHEP07(2024)168",
    journal = "JHEP",
    volume = "07",
    pages = "168",
    year = "2024"
}

@article{Bolton:2024egx,
    author = "Bolton, Patrick D. and Fajfer, Svjetlana and Kamenik, Jernej F. and Novoa-Brunet, Mart{\'\i}n",
    title = "{Signatures of light new particles in B{\textrightarrow}K(*)Emiss}",
    eprint = "2403.13887",
    archivePrefix = "arXiv",
    primaryClass = "hep-ph",
    doi = "10.1103/PhysRevD.110.055001",
    journal = "Phys. Rev. D",
    volume = "110",
    number = "5",
    pages = "055001",
    year = "2024",
    note = "[Erratum: Phys.Rev.D 111, 039903 (2025)]"
}

@article{Rosauro-Alcaraz:2024mvx,
    author = "Rosauro-Alcaraz, S. and Leal, L. P. S.",
    title = "{Disentangling left and right-handed neutrino effects in $B\rightarrow K^{(*)}\nu \nu $}",
    eprint = "2404.17440",
    archivePrefix = "arXiv",
    primaryClass = "hep-ph",
    doi = "10.1140/epjc/s10052-024-13104-x",
    journal = "Eur. Phys. J. C",
    volume = "84",
    number = "8",
    pages = "795",
    year = "2024"
}

@article{Kim:2024tsm,
    author = "Kim, C. S. and Sahoo, Dibyakrupa and Vishnudath, K. N.",
    title = "{Searching for signatures of new physics in $\varvec{B \rightarrow K \, \nu \, \overline{\nu }}$ to distinguish between Dirac and Majorana neutrinos}",
    eprint = "2405.17341",
    archivePrefix = "arXiv",
    primaryClass = "hep-ph",
    doi = "10.1140/epjc/s10052-024-13262-y",
    journal = "Eur. Phys. J. C",
    volume = "84",
    number = "9",
    pages = "882",
    year = "2024"
}

@article{Hati:2024ppg,
    author = "Hati, Chandan and Leite, Julio and Nath, Newton and Valle, Jos{\'e} W. F.",
    title = "{QCD axion, color-mediated neutrino masses, and B+{\textrightarrow}K++Emiss anomaly}",
    eprint = "2408.00060",
    archivePrefix = "arXiv",
    primaryClass = "hep-ph",
    doi = "10.1103/PhysRevD.111.015038",
    journal = "Phys. Rev. D",
    volume = "111",
    number = "1",
    pages = "015038",
    year = "2025"
}

@article{Buras:2024ewl,
    author = "Buras, Andrzej J. and Harz, Julia and Mojahed, Martin A.",
    title = "{Disentangling new physics in $ K\to \pi \nu \overline{\nu} $ and $ B\to K\left({K}^{\ast}\right)\nu \overline{\nu} $ observables}",
    eprint = "2405.06742",
    archivePrefix = "arXiv",
    primaryClass = "hep-ph",
    reportNumber = "MITP-24-049, AJB-24-1",
    doi = "10.1007/JHEP10(2024)087",
    journal = "JHEP",
    volume = "10",
    pages = "087",
    year = "2024"
}

@article{Altmannshofer:2024kxb,
    author = "Altmannshofer, Wolfgang and Roy, Shibasis",
    title = "{Joint explanation of the B{\textrightarrow}{\ensuremath{\pi}}K puzzle and the B{\textrightarrow}K{\ensuremath{\nu}}{\ensuremath{\nu}}{\textasciimacron} excess}",
    eprint = "2411.06592",
    archivePrefix = "arXiv",
    primaryClass = "hep-ph",
    doi = "10.1103/PhysRevD.111.075029",
    journal = "Phys. Rev. D",
    volume = "111",
    number = "7",
    pages = "075029",
    year = "2025"
}

@article{Hu:2024mgf,
    author = "Hu, Quan-Yi",
    title = "{Are the new particles heavy or light in $b \rightarrow s E_{\textrm{miss}}$?}",
    eprint = "2412.19084",
    archivePrefix = "arXiv",
    primaryClass = "hep-ph",
    doi = "10.1140/epjc/s10052-025-14290-y",
    journal = "Eur. Phys. J. C",
    volume = "85",
    number = "5",
    pages = "556",
    year = "2025"
}

@article{Altmannshofer:2025eor,
    author = "Altmannshofer, Wolfgang and Gadam, Sri Aditya and Toner, Kevin",
    title = "{New strategies for new physics search with {\ensuremath{\Lambda}}b{\textrightarrow}{\ensuremath{\Lambda}}{\ensuremath{\nu}}{\ensuremath{\nu}}{\textasciimacron} decays}",
    eprint = "2501.10652",
    archivePrefix = "arXiv",
    primaryClass = "hep-ph",
    doi = "10.1103/PhysRevD.111.075005",
    journal = "Phys. Rev. D",
    volume = "111",
    number = "7",
    pages = "075005",
    year = "2025"
}

@article{Calibbi:2025rpx,
    author = "Calibbi, Lorenzo and Li, Tong and Mukherjee, Lopamudra and Schmidt, Michael A.",
    title = "{Is dark matter the origin of the B{\textrightarrow}K{\ensuremath{\nu}}{\ensuremath{\nu}}{\textasciimacron} excess at Belle II?}",
    eprint = "2502.04900",
    archivePrefix = "arXiv",
    primaryClass = "hep-ph",
    doi = "10.1103/r2gw-rwzw",
    journal = "Phys. Rev. D",
    volume = "112",
    number = "7",
    pages = "075020",
    year = "2025"
}

@article{Lee:2025jky,
    author = "Lee, Jong-Phil",
    title = "{New physics effects in $R(K^{(*)})$, $B_s\toμ^+μ^-$, and $B^+\to K^+ν{\barν}$}",
    eprint = "2502.06370",
    archivePrefix = "arXiv",
    primaryClass = "hep-ph",
    month = "2",
    year = "2025"
}

@article{He:2025jfc,
    author = "He, Xiao-Gang and Ma, Xiao-Dong and Tandean, Jusak and Valencia, German",
    title = "{B {\textrightarrow} K+ invisible, dark matter, and CP violation in hyperon decays}",
    eprint = "2502.09603",
    archivePrefix = "arXiv",
    primaryClass = "hep-ph",
    doi = "10.1007/JHEP07(2025)078",
    journal = "JHEP",
    volume = "07",
    pages = "078",
    year = "2025"
}

@article{Berezhnoy:2025tiw,
    author = "Berezhnoy, Alexander and Lucha, Wolfgang and Melikhov, Dmitri",
    title = "{Analysis of qrec2-distribution for B{\textrightarrow}KMX and B{\textrightarrow}K*MX decays in a scalar-mediator dark-matter scenario}",
    eprint = "2502.14313",
    archivePrefix = "arXiv",
    primaryClass = "hep-ph",
    doi = "10.1103/PhysRevD.111.075035",
    journal = "Phys. Rev. D",
    volume = "111",
    number = "7",
    pages = "075035",
    year = "2025"
}

@article{Bolton:2025fsq,
    author = "Bolton, Patrick D. and Fajfer, Svjetlana and Kamenik, Jernej F. and Novoa-Brunet, Mart{\'\i}n",
    title = "{Impact of new invisible particles on B{\textrightarrow}K(*)Emiss observables}",
    eprint = "2503.19025",
    archivePrefix = "arXiv",
    primaryClass = "hep-ph",
    doi = "10.1103/9rrv-ft75",
    journal = "Phys. Rev. D",
    volume = "112",
    number = "3",
    pages = "035010",
    year = "2025"
}

@article{Aliev:2025hyp,
    author = "Aliev, T. M. and Elpe, A. and Selbuz, L. and Turan, I.",
    title = "{Explaining Belle data on B{\textrightarrow}K(*){\ensuremath{\nu}}{\ensuremath{\nu}}{\textasciimacron} decays via dark Z resonances}",
    eprint = "2503.22347",
    archivePrefix = "arXiv",
    primaryClass = "hep-ph",
    doi = "10.1103/6j6r-9vsl",
    journal = "Phys. Rev. D",
    volume = "112",
    number = "1",
    pages = "015025",
    year = "2025"
}

@article{Chen:2025npb,
    author = "Chen, Chuan-Hung and Chiang, Cheng-Wei and de la Vega, Leon M. G.",
    title = "{Leptoquark-mediated Dirac neutrino mass and its impact on $ B\to K\nu \overline{\nu} $ and $ K\to \pi \nu \overline{\nu} $ decays}",
    eprint = "2503.22431",
    archivePrefix = "arXiv",
    primaryClass = "hep-ph",
    doi = "10.1007/JHEP09(2025)055",
    journal = "JHEP",
    volume = "09",
    pages = "055",
    year = "2025"
}

@article{Ding:2025eqq,
    author = "Ding, Kewen and Li, Ying and Liu, Xuewen and Liu, Yu and Lu, Chih-Ting and Zhu, Bin",
    title = "{Resonant ALP-portal dark matter annihilation as a solution to the B{\ensuremath{\pm}}{\textrightarrow}K{\ensuremath{\pm}}{\ensuremath{\nu}}{\ensuremath{\nu}}{\textasciimacron} excess}",
    eprint = "2504.00383",
    archivePrefix = "arXiv",
    primaryClass = "hep-ph",
    doi = "10.1103/47v4-bn2g",
    journal = "Phys. Rev. D",
    volume = "112",
    number = "11",
    pages = "115034",
    year = "2025"
}

@article{He:2025sao,
    author = "He, Xiao-Gang and Ma, Xiao-Dong and Tandean, Jusak and Valencia, German",
    title = "{Light dark-matter window constrained by K+{\textrightarrow}{\ensuremath{\pi}}++E}",
    eprint = "2505.02031",
    archivePrefix = "arXiv",
    primaryClass = "hep-ph",
    doi = "10.1103/qppt-g39h",
    journal = "Phys. Rev. D",
    volume = "112",
    number = "5",
    pages = "055025",
    year = "2025"
}

@article{DiLuzio:2025qkc,
    author = "Di Luzio, Luca and Nardecchia, Marco and Toni, Claudio",
    title = "{Gauged {\ensuremath{\tau}}-lepton chiral currents and B{\textrightarrow}K(*)Emiss}",
    eprint = "2505.11499",
    archivePrefix = "arXiv",
    primaryClass = "hep-ph",
    doi = "10.1103/7zhp-g199",
    journal = "Phys. Rev. D",
    volume = "112",
    number = "5",
    pages = "055031",
    year = "2025"
}

@article{Shaw:2025ays,
    author = "Shaw, Avirup",
    title = "{Belle II constraints on the nonminimal universal extra dimensional model}",
    eprint = "2509.20027",
    archivePrefix = "arXiv",
    primaryClass = "hep-ph",
    doi = "10.1103/8h4g-rs72",
    journal = "Phys. Rev. D",
    volume = "113",
    number = "9",
    pages = "095027",
    year = "2026"
}

@article{Berezhnoy:2025osn,
    author = "Berezhnoy, Alexander and Lucha, Wolfgang and Melikhov, Dmitri",
    title = "{Scrutinizing Dark-Matter Scenarios with B {\textrightarrow} (K,){\ensuremath{\nu}} Decays}",
    eprint = "2509.23210",
    archivePrefix = "arXiv",
    primaryClass = "hep-ph",
    doi = "10.3390/universe11120385",
    journal = "Universe",
    volume = "11",
    number = "12",
    pages = "385",
    year = "2025"
}

\end{document}